\documentclass[twocolumn,secnumarabic,amssymb, nobibnotes, aps, prd]{revtex4-1}
\usepackage{graphicx}%
\usepackage{siunitx}

\begin{document}

\title{High-angle Deflection of the Energetic Electrons by a Voluminous Magnetic Structure in Near-normal Intense Laser-plasma Interactions}%

\author{J. Peebles$^{1,6}$, A. V. Arefiev$^{1,2}$, S. Zhang$^1$, C. McGuffey$^1$, M. Spinks$^2$, J. Gordon$^2$, E. W. Gaul$^2$, G. Dyer$^{2,5}$, M. Martinez$^2$, M. E. Donovan$^2$, T. Ditmire$^2$, J. Park$^3$, H. Chen$^3$, H. S. McLean$^3$, M. S. Wei$^4$, S. I. Krasheninnikov$^1$ and F. N. Beg$^1$}%
\affiliation{$\mathit{^1}$Department of Mechanical Engineering and Center for Energy Research, University of California San Diego, 9500 Gilman Drive, La Jolla, CA 92093, USA}
\affiliation{$\mathit{^2}$Center for High Energy Density Science, The University of Texas, Austin, 2500 Speedway, TX 78712, USA}
\affiliation{$\mathit{^3}$Lawrence Livermore National Laboratory, 7000 East Avenue, Livermore, CA 94550, USA}
\affiliation{$\mathit{^4}$General Atomics, 3550 General Atomics Court, San Diego, CA 92121, USA}
\affiliation{$\mathit{^5}$SLAC National Acceleration Laboratory, 2575 Sand Hill Rd, Menlo Park, CA 94025}
\affiliation{$\mathit{^6}$Laboratory for Laser Energetics, University of Rochester, 250 E. River Road, Rochester, NY 14623, USA}

\date{\today}

\begin{abstract}
The physics governing electron acceleration by a relativistically intense laser are not confined to the critical density surface, they also pervade the sub-critical plasma in front of the target. Here, particles can gain many times the ponderomotive energy from the overlying laser, and strong fields can grow. Experiments using a high contrast laser and a prescribed laser pre-pulse demonstrate that development of the pre-plasma has an unexpectedly strong effect on the most energetic, super-ponderomotive electrons. Presented 2D particle-in-cell simulations reveal how strong, voluminous magnetic structures that evolve in the pre-plasma impact high energy electrons more significantly than low energy ones for longer pulse durations and how the common practice of tilting the target to a modest incidence angle can be enough to initiate strong deflection. The implications are that multiple angular spectral measurements are necessary to prevent misleading conclusions from past and future experiments.
\end{abstract}

\pacs{numbers!}% insert suggested PACS numbers in braces on next line

\maketitle
\section{Introduction}
The current generation of short pulse lasers reach intensities beyond $\mathrm{I} > 10^{20}\ \mathrm{W/cm^2}$, and hold great potential for many applications in high energy density science such as high energy x-ray backlighters and particle beams\cite{Note1}\cite{Note2}. While lower intensity lasers couple their energy into the transverse motion of electrons, relativistic electrons can couple energy into longitudinal motion via the laser's magnetic field. However, as intensity increases on relativistic laser platforms, the long ns pre-pulse pedestal that arrives prior to the main interaction intensifies as well. This pre-pulse, generated by amplified spontaneous emission processes in the laser, typically has an intensity contrast ratio $\mathrm{(I_{Main Beam}/I_{Pre-Pulse})}$ of $10^{6}$. In high intensity interactions this results in a pre-pulse of high enough intensity to ionize material, creating an underdense pre-plasma that extends for hundreds of\SI{}{\ \micro\meter}. 

While pre-plasma has been seen as detrimental to electron acceleration for applications such as fast ignition due to filamentation and self focusing instabilities\cite{MacPhee}\cite{Sunahara}, it actually can be beneficial in other ways, enabling a number of mechanisms that can generate energetic electrons. Accelerating electrons is the primary way to couple the energy in a laser to a plasma and provides the basis for a wide range of phenomena and applications. The generation of large quantities of x-ray and energetic secondary particles, such as ions \cite{Pomerantz}, neutrons\cite{Schollmeier} and positrons\cite{Chen} highly depend on relativistic electrons. The electron acceleration in these experiments occur in a regime where the plasma response time is short compared to the laser pulse. In these experiments the density profile evolves slowly relative to the laser and develops into a quasi-steady state, opposite of the regime necessary for processes such as wakefield acceleration. Therefore, understanding the parameters that control the quantity, energy and trajectory of these electrons is critical to properly understanding the production of other energetic particles from secondary interactions.

A common element for electron acceleration mechanisms is the important role played by quasi-static transverse and longitudinal pre-plasma electric fields in enhancing the energy transfer from the laser pulse to pre-plasma electrons \cite{Booshan1}-\cite{Robinson}. These fields are relatively weak compared to the field of the laser pulse and are unable to directly transfer considerable energy to the electrons. However, they do change the phase between the oscillating electric field of the laser and the electron velocity, which can result in a net energy gain with each laser period. This is the essence of the mechanism called direct laser acceleration (DLA) that leads to acceleration of the so-called super-ponderomotive electrons in an extended pre-plasma. These electrons have a corresponding relativistic $\gamma$-factor greatly exceeding the conventional estimate of $\gamma \approx a_0^2$, where $a_0$ is the normalized laser amplitude. Indeed it has been shown that these super-ponderomotive electrons can enhance the energy of target normal sheath accelerated (TNSA) protons for radiography purposes but requires carefully controlled conditions for the rise time and trajectory of the electron beam \cite{Kim}. 

Prior experiments using pulses with durations of 150 fs \cite{Peebles},400 fs \cite{Tanaka}, 500 fs \cite{Toshi} and 700 fs \cite{Ma} suggest that appreciable quantities of super-ponderomotive electrons should be accelerated for pulse lengths 400 fs or longer. The goal of the experiment presented here was to fill a gap between theory and experiment by accelerating large quantities of super-ponderomotive electrons using a very high intensity Texas Petawatt (TPW) laser while varying the interaction pulse length. These super-ponderomotive electrons were observed with energy up to $150$ MeV or more. However, contrary to prior experiments and simulations they were only detected at the longest pulse length tested (600 fs). In this paper we report the experimental measurement of super-ponderomotive electrons accelerated by DLA and the significance that a quasi static magnetic field has on their trajectory.

\section{Experimental Setup and Initial Simulations}

\begin{figure}[b]
	\includegraphics[width=\columnwidth]{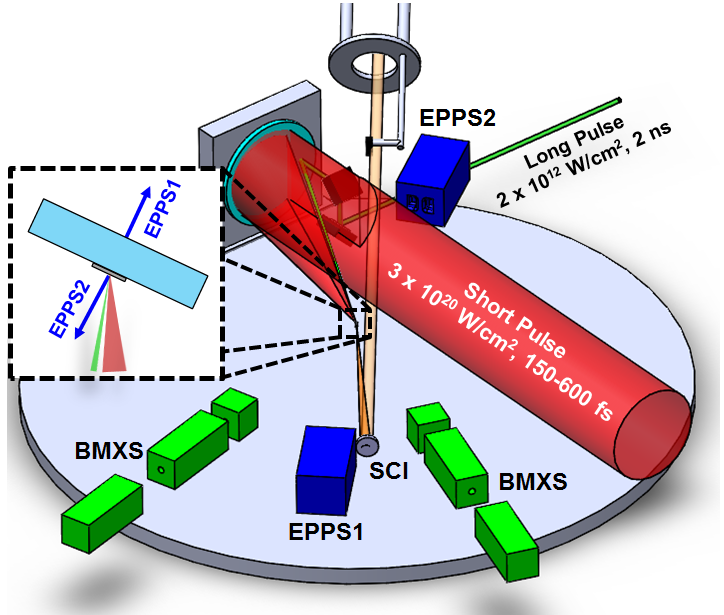}
	\caption{Setup for the experiment on the TPW laser. A long pulse beam (green) generates an underdense plasma on the surface of the target while the high intensity short pulse (red) accelerates electrons measured by the primary EPPS diagnostics.}
\end{figure}

Similar to previous experiments \cite{Peebles}-\cite{Willingale} a low intensity, 2 ns beam with a large focal spot was used to generate a controllable, uniform, underdense plasma. This beam was injected 4 ns prior to the arrival of a short pulse beam, which had a variable pulse length (150, 450, 600 fs) and energy (30-105 J). Intensity was kept nominally to $2-3 \times 10^{20}\ \mathrm{W/cm^2}$ for all pulse lengths, and the beam was focused to a \SI{10}{\ \micro\meter} diameter spot, incident on the target at a $21.8^{\circ}$ angle. The intrinsic pre-pulse from amplified spontaneous emission of the high intensity beam was nearly negligible after recent upgrades to the TPW, which improved the pre-pulse intensity contrast to over $3 \times 10^{10}$ \cite{GaulContrast}. Both beams had a wavelength of 1057 nm. The high-contrast laser was incident on a 1 $\mathrm{mm^2}$ \SI{100}{\ \micro\meter} thick planar foil target composed primarily of aluminum, with a buried \SI{20}{\ \micro\meter} thick copper layer for diagnostic purposes. This foil was attached to a bulk plastic block used to inhibit electron refluxing, preventing double counting of electrons in our diagnostics. The experimental setup with the positions of diagnostics are shown in Figure 1.

\begin{figure}[b]
	\includegraphics[width=\columnwidth]{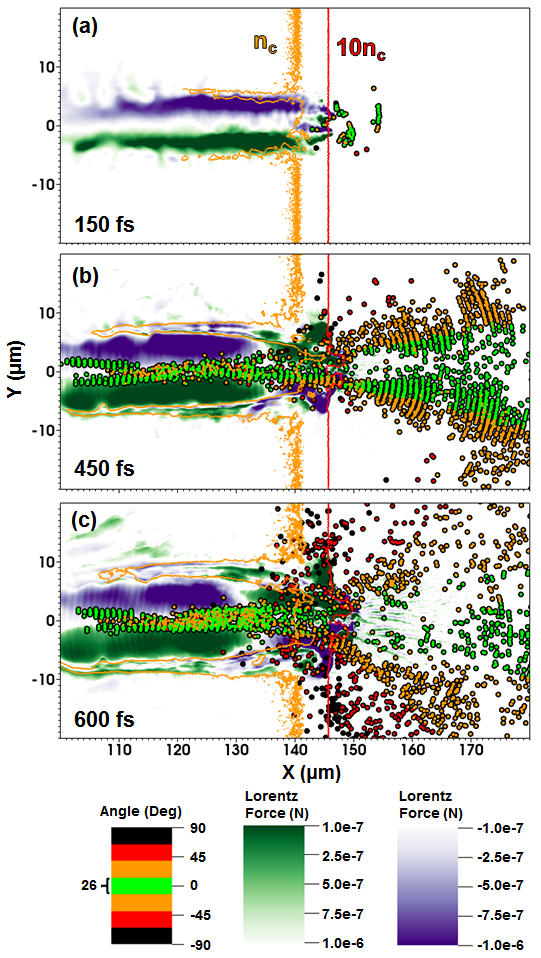}
	\caption{Simulation results of a 150, 450 and 600 fs beam (traveling left to right) incident on a target with underdense plasma. Electrons with energy $>$ 100 MeV traveling in the forward direction are color coded by angle relative to the target normal. The deep purple and green sections represent areas with significant electromotive force in the transverse direction on a 100 MeV electron.}
\end{figure}

Diagnostics were fielded to characterize hot electron temperature and trajectory. While bremsstrahlung spectrometers (BMXS) and copper K$\alpha$ imaging (SCI) were used, their signals are more sensitive to lower temperature electrons. High energy electrons were measured with EPPSes, a set of  calibrated magnetic electron spectrometers \cite{EPPS}. The spectrometers had 1 x 2 mm entrance pinholes, a strong magnet to deflect charged particles and differentiate them by energy and charge and two image plate detectors. The energy range was 5-100 MeV with energy resolution scaling with energy, from 5 keV at low energies to 1 MeV at higher energies. Two spectrometers were placed in the chamber, the first 43 cm away from the interaction facing the rear surface normal of the target and the second was placed 55.5 cm away from the interaction, 3 degrees off the front surface normal of the target. Spectra taken from the EPPS were characterized by the Half Maximum Integrated Energy (HMIE) value, which represents the energy value where 50 percent of the total energy in the spectrum is contained, e.g. a HMIE of 10 MeV means 50\% of the energy in the spectrum is contained in electrons with less than 10 MeV energy. Hence a higher HMIE value will result from a spectrum with a greater proportion of high energy electrons.

Initial hydrodynamic simulations with 2D FLASH \cite{FLASH} were conducted to estimate the density profile of the underdense plasma using the parameters of the long pulse beam in the experiment. The resulting density profile was approximated as a sum of exponential decays with a density of around $10^{19}\ \mathrm{cm}^{-3}$ \SI{150}{\ \micro\meter} away from the solid surface. This density profile was input into the 2D particle-in-cell (PIC) code EPOCH \cite{EPOCH} and truncated \SI{160}{\ \micro\meter} away from the target surface. The simulation box size was \SI{250}{\ \micro\meter} long, \SI{60}{\ \micro\meter} wide, with 10 electron and 5 proton macro particles per cell. The resolution was 30 cells/\SI{}{\\\micro\meter} in the x direction and 15 cells/\SI{}{\\\micro\meter} in the y direction. A laser was injected into the box with an intensity of $3 \times 10^{20}\ \mathrm{W/cm^2}$, a 70 fs rise time and a \SI{10}{\ \micro\meter} diameter spot.

In Figure 2 we present initial snapshots simulations for pulse lengths of 150, 450 and 600 fs taken just after the center of the pulse impacts the target in time. With the 150 fs pulse no significant super-ponderomotive electrons were accelerated. For 450 fs super-ponderomotive electrons are accelerated and travel primarily along the laser trajectory, while at 600 fs these electrons are dispersed by fields that develop near the critical density. Based on these simulations we anticipated measurable quantities of super-ponderomotive electrons ($\mathcal{E} >$  60 MeV, $a_0$ = 15.6) using pulse lengths greater than 400 fs.

\section{Results and Discussion}

Over 35 experimental shots were performed over the course of the campaign with time split evenly between the 3 pulse length settings (150, 450 and 600 fs); half of the total shots did not use the long pulse beam to serve as the no pre-plasma cases. We verified that no super-ponderomotive electrons were measured with the shortest pulse length. When comparing the spectra for the 150 fs pulse length case there was no difference in the HMIE values when the long pulse beam was included. Upon increasing the pulse length to 450 fs the results were largely the same and no significant departure was seen between shots with and without pre-plasma. This was a surprising result since prior experiments \cite{Tanaka}\cite{Toshi} and our simulations with similar pulse lengths exhibited super-ponderomotive electrons and a change in spectrum shape.

\begin{figure}[b]
	\includegraphics[width=\columnwidth]{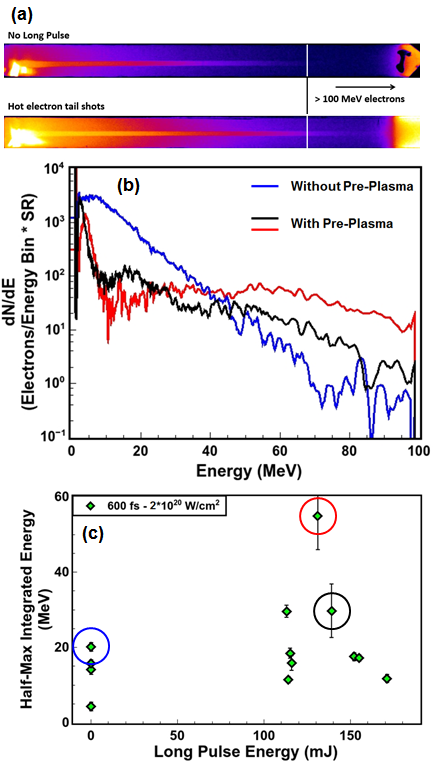}
	\caption{(a): Raw data from EPPS1 comparing cases with and without underdense plasma. Background subtracted lineouts are taken of the data and convolved with the spectrometer's calibrated dispersion to produce electron spectra. \\\hspace{\textwidth} (b): Electron spectra leaving the rear of the target measured by EPPS1 in the calibrated region of the diagnostic. Two 600 fs shots with pre-plasma shown in black and red produced substantial quantities of super-ponderomotive electrons $>$ 60 MeV resulting in a dramatic change to the electron spectrum compared to the no pre-plasma case (blue). \\\hspace{\textwidth}(c): HMIE values taken for all shots at 600 fs. Shots from the above spectra are circled with their respective color. At 600 fs, 3 out of 9 shots with underdense plasma measured a significant quantity of super-ponderomotive electrons}
\end{figure}

Extending the pulse length further to 600 fs yielded significant changes to the electron spectrum shown in Figure 3; 3 out of the 9 shots with the long pulse beam had differently shaped spectra than their counterparts. These spectra contained electrons so energetic that they were detected at the end of the spectrometer (beyond its known calibration) and had energy exceeding 150 MeV. These results contradict the initial expectations from simulations shown in Figure 2. At 450 fs in simulations copious quantities of high energy electrons are accelerated nearly \SI{50}{\ \micro\meter} from the target surface and move forward into the target, while in the experiment no super-ponderomotive electrons were measured. At 600 fs super-ponderomotive electrons were measured on experiment while in simulations most electrons were severely deflected and few super-ponderomotive electrons propagated directly forward toward the diagnostic.

\begin{figure}[t]
	\includegraphics[width=\columnwidth]{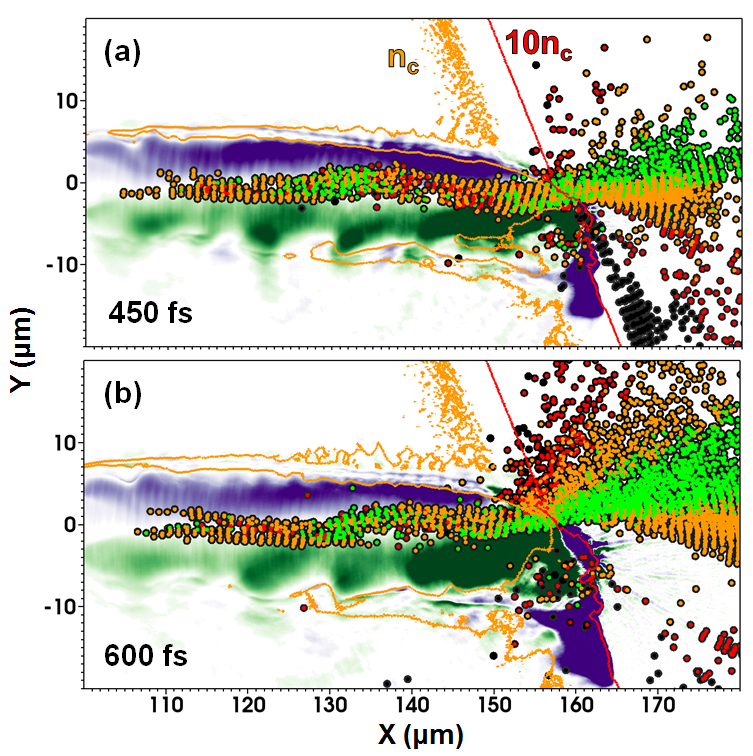}
	\caption{Identical simulation setup to Figure 2 with the laser at a 20 degree incidence angle on target. The electrons are color coded by angle bin with respect to the EPPS diagnostic placed in the target normal direction. At 450 fs electrons travel in the direction of the laser similar to those in Figure 2 (b), while at 600 fs the electrons are deflected upwards in the target normal direction. This is due to the development of a large asymmetry in the electromotive force.}
\end{figure}

What could cause this large discrepancy between the simulations and experimental data? The cause can be partly attributed to the EPPS diagnostic itself. The EPPS collects electrons only near the target rear normal and the solid angle ($\sim 10^{-5}$ steradian) is an extremely small fraction of the solid angle where electrons are seen to travel in the simulations. Another difference between simulation and experiment lies in the inclusion of the laser incidence angle. In our experiment the laser was incident at roughly a $\mathrm{20^o}$ angle and the EPPS diagnostic was placed facing the target normal, while our simulations used a normal incidence beam. When conducting another simulation with the same parameters but with the laser incidence angle and diagnostic angles included, the trajectories of super-ponderomotive electrons were altered significantly.

Figure 4 shows the results from the angled simulation taken at the same times as those in Figure 2. Initially the $>$ 100 MeV electrons propagate primarily in the laser direction, not in the direction of the diagnostic. Measuring the angle of the electrons with respect to the viewing angle of the EPPS shows that the EPPS would miss the majority of super-ponderomotive electrons at 450 fs. However, later in time, electrons were scattered weakly toward the target normal rather than in all directions. At 600 fs the bulk of high energy electrons were deflected in the direction of the target normal and EPPS diagnostic, resulting in a greater chance of detection, which is directly supported by the experimental results.

\begin{figure}[t]
	\includegraphics[width=\columnwidth]{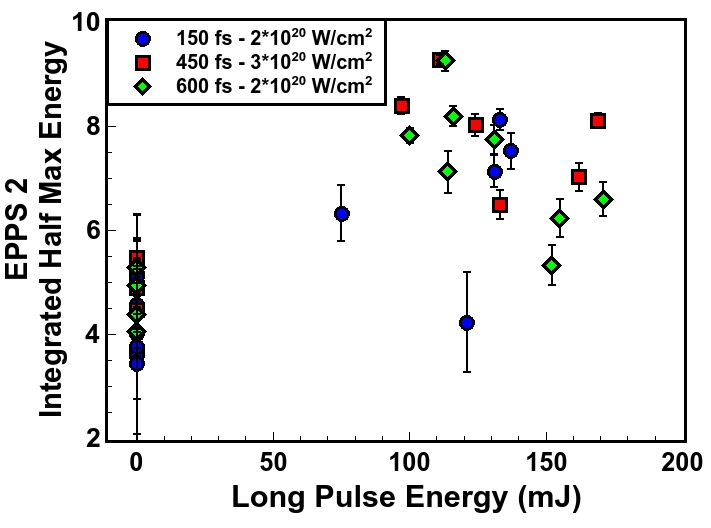}
	\caption{HMIE values for electrons measured from the front surface of the target. Higher energy electrons were measured when an underdense plasma was present, regardless of pulse length suggesting that electrons in the pre-plasma were accelerated by the reflected beam.}
\end{figure}
\begin{figure}[]
	\includegraphics[width=\columnwidth]{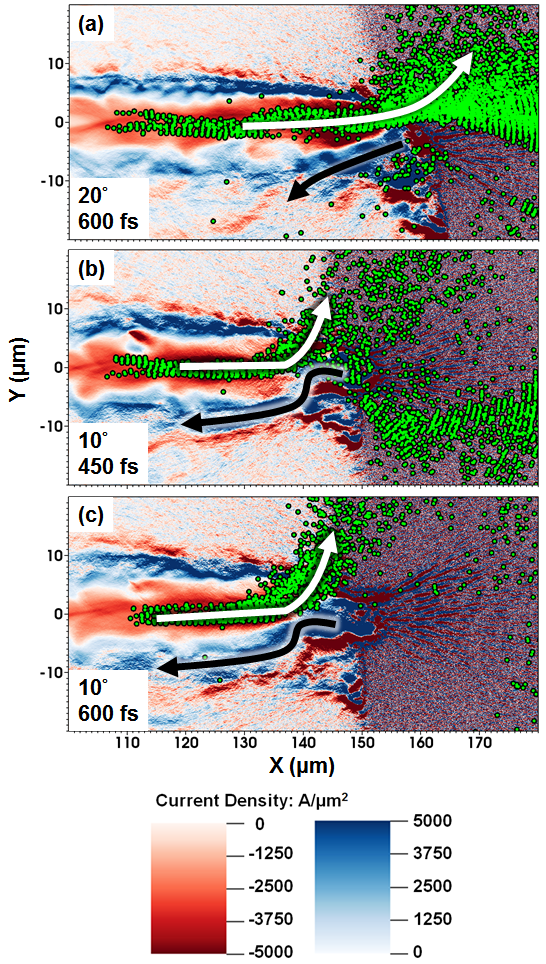}
	\caption{(a): A current density plot for the 20 degree laser incidence angle case taken at the same time as Figure 4 (c). The effect of the asymmetric magnetic field on incident hot electrons is gradual causing a smaller deflection. Arrows indicate the location and direction of the majority of current flow.\\\hspace{\textwidth} (b): A plot of a simulation with a 10 degree laser incidence angle taken earlier in time. \\\hspace{\textwidth} (c): Plot of the 10 degree incidence simulation at the same time as the top 20 degree image. The current accelerated by the reflected beam expands and encounters the incident current more directly than in the 20 degree case. The super-ponderomotive electrons see a sudden, larger increase in the deflecting magnetic field as the fields join together.}
\end{figure}

Upon closer examination of electron deflection at these later times it is found that the electrons encounter a counter-propagating current of colder electrons. The effect of this reverse current can be seen by mapping the electromagnetic force on a forward going electron (seen in purple and green in Figs 2 and 4). The forward going electrons generate a self confining magnetic field based on their current. Extending from the target surface, a counter-propagating current generates an opposite field that causes the electrons to split upwards or downwards. As pulse length increases this opposing field gains strength and causes more severe deflections. This effect has the largest impact on super-ponderomotive electrons, as they are accelerated via DLA far from the target surface and travel through this field. Lower energy, ponderomotive scaling electrons are less affected since they are accelerated near the critical density, closer to the target surface and do not interact with these fields over an extended distance.

The development of counter-propagating current can be traced to electrons in the pre-plasma near the target surface that are accelerated by reflected laser light away from the critical surface. This effect has been characterized in 3D PIC simulations conducted by F. Perez \textit{et al.}\cite{Perez}, which show the development of strong magnetic fields due to counter-propagating current after 473 fs. In the normal incidence case the two currents directly oppose each other and incoming electrons are deflected in all directions; in the angled case the opposing fields develop at an angle with respect to the target surface. This angle causes the fields from the counter-propagating current to build upon the fields from the incident current rather than oppose them, leading to a large asymmetry and significant deflection toward target normal.

Experimental evidence of pre-plasma electrons accelerated by reflected light was found in the data of the spectrometer facing the front surface of the target (EPPS2). Measuring the HMIE values for electrons traveling in this direction shows distinctly higher energies for nearly all shots with the long pulse beam, regardless of pulse length (Fig 5). However, as our simulations show, the field from this reflected current requires time to develop and only has enough strength to deflect significantly electrons after 500 fs.

\section{Summary and Future Considerations}

In summary, we found that in experiments no super-ponderomotive electrons were detected in the forward direction for $\tau_L$ = 150 fs, nor for $\tau_L$ = 450 fs, in contradiction to earlier normal-incidence simulations, and only sporadically for $\tau_L=$ 600 fs. Simulations including the experimental incidence angle showed that the super-ponderomotive electrons in fact were generated as early as $450$ fs, but were likely traveling in the direction of the laser direction and not towards the diagnostic placed in the target normal direction. As fields from an opposing current increased later in time these electrons were eventually deflected preferentially in the direction of the diagnostic. These results raise the question, how sensitive are super-ponderomotive electron trajectories to changes in laser incidence angle? To address this, another simulation with identical parameters was conducted, which used a 10 degree incidence angle rather than 20. As seen in figure 6 (c), electrons in the small angle case were deflected \textit{more} significantly than in either of the previous two cases (0 and 20 degree), with a large portion deflected away from the target entirely late in time. Due to the smaller incidence angle, the region where laser light reflects off the critical density surface becomes narrower causing the opposing current to become more concentrated. More importantly, halving the incidence angle causes the relative angle between the reflected and incident light to be quartered. The opposing current therefore has a more head on trajectory with the incident electrons leading to greater deflection angle unlike the ``glancing blow'' seen in the large angle case.

These results point to a few considerations for future experiments and simulations. Understanding the effect of laser incidence angle is crucial for setting up an experiment. Many high intensity laser facilities do not allow for direct normal laser incidence due to the potential damage caused by reflected light, which means that incidence angle effects are a factor that must be considered for experiments that have been or will be performed at these facilities. Even for experiments that use perfectly normal incidence, the effect of counter-propagating current can result in deflecting electrons away from the laser direction as shown in Figure 2. Electron trajectories change significantly when incidence angle is varied and must be taken into consideration when placing diagnostics and when accelerating protons via TNSA using non-relativistic beams \cite{Kim}. Due to the nature of high energy electrons, diagnostics capable of measuring super-ponderomotive electrons use very small solid angle apertures. Measuring these directional, high energy electrons poses a challenge to the diagnostics community to develop new tools capable of providing spatial information for higher energy electrons.

%Since these electrons are highly directional, understanding the effect of incidence angle becomes all the more important. Therefore, prior to any high intensity experiment, 2D PIC simulations must be conducted in order to anticipate the trajectory of electrons, since the degree of deflection is highly dependent on the exact experimental parameters. Otherwise simply placing a high energy spectrometer behind the target at an arbitrary angle may fail to detect super-ponderomotive electrons despite their presence in the experiment.

\section*{Acknowledgments}

Experimental time and resources were supported by the United States Department of Energy under contract DE-NA0001858 (DOE). Simulations were performed using the EPOCH code (developed under UK EPSRC Grants No. EP/G054940/1, No. EP/G055165/1, and No. EP/G056803/1) using HPC resources provided by the TACC at the University of Texas. The work by A. Arefiev ,CHEDS researchers, and the TPW itself were supported by NNSA Cooperative Agreement DE-NA0002008.

\end{document}